\documentclass[aps,prb,twocolumn,superscriptaddress,floatfix]{revtex4}
\usepackage{psfig}

\bibstyle{apsrev.bib}

\begin{document}
\input epsf.sty

\title{Nonequilibrium critical dynamics in inhomogeneous systems}

\author{Michel Pleimling}
\affiliation{Institut f\"ur Theoretische Physik I, Universit\"at
Erlangen-N\"urnberg, D-91058 Erlangen, Germany}

\author{Ferenc Igl\'oi}
\affiliation{
Research Institute for Solid State Physics and Optics,
H-1525 Budapest, P.O.Box 49, Hungary}
\affiliation{
Institute of Theoretical Physics,
Szeged University, H-6720 Szeged, Hungary}

\date{\today}

\begin{abstract}
  We study nonequilibrium dynamical properties of inhomogeneous
  systems, in particular at a free surface or at a defect plane. Thereby we
  consider nonconserved (model-A) dynamics of a system which is
  prepared in the high-temperature phase and quenched into the
  critical point. Using Monte Carlo simulations we measure single spin
  relaxation and autocorrelations, as well as manifold
  autocorrelations and persistence. We show that, depending on the
  decay of critical static correlations, the short time dynamics can
  be of two kinds. For slow decay of local correlations the usual
  domain growth process takes place with non-stationary and algebraic
  dynamical correlations. If, however, the local correlations decay
  sufficiently rapidly we have the so called cluster dissolution
  scenario, in which case short time dynamical correlations are
  stationary and have a universal stretched exponential form. This
  latter phenomenon takes place in the surface of the
  three-dimensional Ising model and should be observable in real
  ferromagnets.

\end{abstract}

\maketitle

\newcommand{\bc}{\begin{center}}
\newcommand{\ec}{\end{center}}
\newcommand{\be}{\begin{equation}}
\newcommand{\ee}{\end{equation}}
\newcommand{\beqn}{\begin{eqnarray}}
\newcommand{\eeqn}{\end{eqnarray}}

\section{Introduction}

To characterize the dynamical universality class of a critical system
at equilibrium it is enough to provide the value of the dynamical
exponent, $z$, which generally depends on the local dynamics,
conservation laws and symmetries.\cite{hh77} In a nonequilibrium
system, which is prepared by quenching it from the high-temperature
initial state to the critical point, new nonequilibrium critical
exponents have to be defined.\cite{jss89,huse89} The reason of this is
the broken time translation invariance due to a discontinuity at the
time horizon ("time-surface"). Since at the critical point there is no
characteristic time-scale the nonequilibrium preparation has a
long-lasting effect, which has two important consequences. First, the
dynamical correlations, such as the single spin autocorrelation
function, $C(t,s)$, are generally non-stationary.  They depend on both
the preparation or waiting time, $s$, and the observation time, $t$.
The second effect is the appearance of a new nonequilibrium critical
exponent, $\lambda$, which is used to describe the long-time
asymptotic, $t \gg s$, of the autocorrelation function as $C(t,s) \sim
t^{-\lambda/z}$.\cite{huse89}

Another, and related process of nonequilibrium dynamics is the
relaxation of the magnetization, if in the initial state there is a
small, non-vanishing value of $m_i$. For short times the magnetization
has a power-law dependence: $m(t) \simeq m_i t^{\theta}$, with the
initial slip exponent $\theta$ which satisfies the scaling relation:
$\lambda=d-\theta z$.\cite{jss89,jans92} Here $d$ is the spatial
dimension of the system. In mean-field theory $\theta=0$, whereas in
real system generally $\theta>0$. Thus due to fluctuations the order
is increasing in the early time regime. This phenomenon is related to
the fact that in the initial state there are no long-range
correlations, thus the systems is mean-field like. Since the actual
critical temperature of the system, $T_c$, is lower than the critical
temperature of the mean-field model, $T_c(mf)>T_c$, in the early
time-steps there is an effective coarsening process, during which the
magnetization is increasing. The nonequilibrium magnetization with
$m_i < 1$, however, can not exceed the value obtained with $m_i=1$,
when from the ordered initial state the magnetization decreases as
$m(t) \sim t^{-x/z}$, where $x$ is the anomalous dimension
of the magnetization. This critical parameter is defined trough the
asymptotic decay of the equilibrium, equal-time correlation function
at the critical point: $\langle \sigma_i(t) \sigma_j(t) \rangle \sim |i-j|^{-2x}$ and
satisfies the scaling relation: $x=\beta/\nu$, with $\beta$ and $\nu$
being the usual static
magnetization and correlation length exponents, respectively. From
the previous reasoning we obtain for the border of the short-time regime,
$t_i$, as $t_i \sim m_i^{-z/x_i}$, where $x_i=\theta z +x$ is the
anomalous dimension of the initial magnetization. Note that
the nonequilibrium time-scale, $t_i$,
diverges as $m_i \to 0$.

Besides single spin autocorrelations one often considers the dynamics
of extended objects, such as the order-parameter of the whole sample
(global autocorrelations) or that of a manifold\cite{mb03} with dimension $d'<d$.
In the asymptotic regime $t \gg s$ the manifold autocorrelation
function displays an algebraic decay\cite{mb03}: $G(t) \sim t^{-\lambda'/z}$ with
$\lambda'=\lambda-d'$.  Still another dynamical quantity is
represented by the persistence, $P_{pr}(t)$, which measures the
probability that the order-parameter associated either to a single
spin, or to a manifold, or to the whole sample has not changed sign
within time $t$.\cite{maju99} Among others, persistence is related to the
properties of the given autocorrelation function and for
global and manifold persistence it has often a power-law dependence:
$P_{pr}(t) \sim t^{-\Theta_{pr}}$, where $\Theta_{pr}$ is a new
nonequilibrium exponent.\cite{mbcs96} Only for global persistence and
for a Markovian process can $\Theta_{pr}$ be expressed by other known
exponents.\cite{mbcs96}

The dynamical critical behavior, as outlined above, refers to the bulk
of a homogeneous system in the thermodynamic limit. Real materials,
however, are bounded by surfaces and many nonequilibrium processes
(thermalization, transfer of heat, etc.) are very intensive at the
surface. At a free boundary layer the order is weaker than in the bulk
due to missing bonds and correlations between two surface sites, $i_1,
j_1$: $\langle \sigma_{i_1}(t) \sigma_{j_1}(t) \rangle \sim |i_1-j_1|^{-2x_1}$
involve a new scaling dimension, $x_1>x$.\cite{binder,diehl,ipt,pleim} This satisfies the
scaling relation: $x_1=\beta_1/\nu$, with $\beta_1$ being the surface
magnetization exponent. Note that there is an analogy
between spatial and temporal surfaces, where translational invariance
in space and in time, respectively, is broken. Similarly the
scaling dimensions, $x_1$ and $x_i$, respectively, play analogous
roles. According to
field-theoretical investigations \cite{dd83} no new dynamical critical
exponent should be introduced at the surface: $z$ and $x_i$ remain
unchanged whereas $\theta_1$ and $\lambda_1$, as defined on surface
spins, can again be expressed by the known exponents, including
$x_1$.\cite{RiCz_95,ms96}

We note that besides a free surface there
exist other types of inhomogeneities, such as a localized or an
extended defect plane, near which scaling of the local magnetization
is different from that in the bulk. Correlations between two sites
at the defect plane, $i_l, j_l$ are asymptotically given at the
critical point by $\langle \sigma_{i_l}(t) \sigma_{j_l}(t)
\rangle \sim |i_l-j_l|^{-2x_l}$, which involves a new static (local)
exponent $x_l \ne x$.\cite{ipt} In an inhomogeneous system it can be
found in many cases that $x_l>x_i$, thus the spatial surface or
interface is more disordered than the "temporal surface". This happens
for example at the ordinary surface transition of the
three-dimensional Ising model, with $x_1=1.26$ and $x_i=0.74$, and
should be realized in real magnets, too.  Nonequilibrium dynamical
critical behavior of systems with $x_l>x_i$ is expected to be
different from that with $x_l<x_i$. In the latter case traditional
domain growth (DG) scenario should hold. On the other hand in the
former case critical relaxation is more effective than the
nonequilibrium growth, therefore no new domains are created in the
surface or interface region. The dynamical process is dominated by
rare clusters with correlated sites, and the nonequilibrium dynamics
is realized through cluster dissolution (CD). The CD process can be
seen also for the surface manifold autocorrelation function, provided
$d'-x_l-x<0$.\cite{pi04}

Our aim in the present paper is to study in detail the nonequilibrium
dynamical critical behavior in inhomogeneous systems, in particular at
a free surface or at defect planes. Here we restrict ourselves to
nonconserved (model-A) dynamics.\cite{hh77} We use scaling theory and perform
extensive Monte Carlo simulations. We clarify the properties of the
nonequilibrium dynamics in the DG and CD regimes. A short account of
our results obtained at a free surface has been announced in a
Letter.\cite{pi04}

Our paper is organized as follows. In Sec.\ 2 we start with the
presentation of a phenomenological picture based on scaling theory.
Sec.\ 3 contains our numerical results on local magnetization
relaxation and autocorrelations i) at free surfaces in the two- and
three-dimensional Ising models, ii) at an extended surface defect in
the two-dimensional Ising model (Hilhorst-van Leeuwen (HvL)
model),\cite{hvl} and iii) at an internal defect line in the
two-dimensional Ising model (Bariev model).\cite{Bar79} We also
consider iv) surface manifold autocorrelations and persistence. In
Sec.\ 4 we close our paper by a discussion.

\section{Scaling theory of nonequilibrium dynamics in inhomogeneous systems}

We consider a $d$-dimensional system having a free surface or a defect
plane, the distance of it being denoted by $y$. (In the following we
will refer to that plane as a {\it surface} and the local scaling
dimension will be called $x_l$, so that $x_l=x_1$ at a free surface.)
From the high-temperature phase, $T > T_c$, we quench the system to
its critical temperature and study different physical quantities
(magnetization, single spin autocorrelation, manifold autocorrelation)
as a function of the waiting time, $s$, and the observation time, $t$,
both measured from the time of the quench. We start our analysis with
the relaxation of the magnetization.

\subsection{Relaxation of the magnetization}

In this case the system is prepared with a small initial
magnetization, $m_i > 0$, and the local magnetization $m=\langle
\sigma_y(t) \rangle \equiv m(y,t,m_i)$ is measured at the critical
temperature. Here $\sigma_y(t)$ is the operator of the local
magnetization. Under a scaling transformation, $l \to l/b$, when
length, $l$, is rescaled by a factor, $b>1$, the magnetization behaves
as:
\be
m(y,t,m_i)=b^{-x} m(y/b,t/b^z,m_i b^{x_i})\;.
\label{scal_m}
\ee
Here $x_i$ is the new, nonequilibrium scaling dimension of the initial
magnetization, as introduced in the previous section. Taking in
Eq.(\ref{scal_m}) $b=t^{1/z}$, we arrive to:
\be
m(y,t,m_i)=t^{-x/z} \tilde{m}(y/t^{1/z},m_i t^{x_i/z})\;,
\label{scal_m1}
\ee
in which the scaling function, $\tilde{m}(r,\mu)$, behaves differently
in the bulk, $r \to \infty$, and at the surface, $r \to 0$,
respectively.

\subsubsection{Bulk behavior}

In the bulk of the system we have by definition $\lim_{r \to \infty}
\tilde{m}(r,\mu)=\tilde{m}_b(\mu)$, and the bulk scaling function
$\tilde{m}_b(\mu)$ has a cross-over:

\be
\tilde{m}_b(\mu)=\cases{\sim \mu,\quad \mu \ll 1 \cr
               {\rm const},\quad \mu \gg 1 \cr} \label{m_b}
\ee
For short times, $t<t_i \sim m_i^{-z/x_i}$, the magnetization increases
as $m(t) \sim t^{\theta}$, with $\theta=(x_i-x)/z$. For longer times,
however, the magnetization follows the equilibrium decay, $m(t) \sim
t^{-x/z}$. We note that the short-time behavior of the magnetization
is the result of two different processes. Due to nonequilibrium domain
growth the magnetization increases as $t^{x_i/z}$, whereas critical
relaxation acts to reduce its value by $t^{-x/z}$.

\subsubsection{Surface behavior}

In order to understand the concept of surface critical behavior we start with
the equilibrium magnetization, $m_{\rm eq}(y,t,\delta)$, where $\delta=(T_c-T)/T_c$
is the reduced critical temperature. The scaling transformation now reads as:
\be
m_{\rm eq}(y,t,\delta)=b^{-x} m_{\rm eq}(y/b,t/b^z,\delta b^{1/\nu})\;.
\label{scal_m2}
\ee
We consider first the {\it static} behavior with $t=0$, set the
length-scale to $b=\delta^{-\nu}$ and obtain $m_{\rm
  eq}(y,t=0,\delta)=\delta^{x \nu} \tilde{m}_{\rm eq}(y \delta^{\nu})$.
  Here we notice the temperature dependence of the bulk magnetization:
  $m_{\rm eq}^b \sim \delta^{\beta}$, with a critical exponent
  $\beta=x \nu$ and with the scaling function: $\lim_{\tilde{r} \to \infty}
  \tilde{m}_{\rm eq}(\tilde{r})=const$. By definition the magnetization at the
    surface behaves as: $m_{\rm eq}^1 \sim \delta^{\beta_l}$ with
    $\beta_l=x_l \nu$. This is compatible with the above scaling
    result provided the scaling function for a small argument (short
    distance) behaves as: $\lim_{\tilde{r} \to 0} \tilde{m}_{\rm eq}(\tilde{r}) \sim
      \tilde{r}^{x_l-x}$.  Now considering the equilibrium {\it dynamical
        behavior} at the critical point we set in Eq.(\ref{scal_m2})
      $\delta=0$ and $b=t^{1/z}$. Repeating the previous reasoning we
      obtain for the magnetization in the surface region $m_{\rm
        eq}(y,t)=t^{-x/z}\tilde{m}_{\rm eq}(r) \sim t^{-x/z}
      r^{x_l-x}$, $r \ll 1$, which can be obtained from the
      short-distance expansion in field-theoretical
      calculations.\cite{diehl} Note that starting from the bulk
      order is decreasing towards the surface for $x_l>x$. In
      nonequilibrium relaxation two processes have to be taken into
      account close to a surface: the magnetization i) increases due
      to nonequilibrium domain growth and ii) decreases due to surface
      effects. Of course, the first process is of no importance when
      starting from a fully ordered state.\cite{Kik85,Rie85} The
      relative strength of the two processes are given by the scaling
      dimensions $x_i$ and $x_l$, respectively. Nonequilibrium
      dynamics is markedly different for $x_i > x_l$, which is the
      domain growth regime, and for $x_i < x_l$, which is the regime
      of cluster dissolution.

\paragraph{Domain growth: $x_i > x_l$}

In this case bulk nonequilibrium order penetrates into the surface
region, in which, as time goes on, new domains are formed. Therefore
short distance expansion of the magnetization stays valid, yielding
for small $r$ the following expression for the scaling function
$\tilde{m}(r,\mu)$ given in Eq.(\ref{scal_m1}):
\be
\lim_{r \to 0}\tilde{m}(r,\mu)=r^{x_l-x} \tilde{m}_l(\mu), \quad x_i>x_l \;.
\label{m_1}
\ee
Here $\tilde{m}_l(\mu)$ has the same limiting $\mu$-dependence as
$\tilde{m}_b(\mu)$ in Eq.(\ref{m_b}). Thus we have in the short-time
limit, $t<t_i \sim m_i^{-z/x_i}$, $m_l(t) \sim t^{\theta_l}$, with
$\theta_l=\theta-(x_l-x)/z=(x_i-x_l)/z$. Thus the initial slip
exponent at the surface in the DG regime can be expressed by known
exponents.\cite{RiCz_95}

\paragraph{Cluster dissolution: $x_i < x_l$}

In this case, for short times nonequilibrium order can not penetrate
into the surface region and therefore no new domains are formed there.
In contrary the initially existing ordered clusters are diluted and
the initial relaxation has a fast, non-algebraic dependence. Evidently
the short distance expansion is not valid here. Analyzing the CD
process we have determined in the Appendix the small $r$ behavior of
the scaling function, $\tilde{m}(r,\mu)$. It is given in a stretched
exponential form:
\be
\lim_{r \to 0}\tilde{m}(r,\mu) \sim \exp(-C r^{\kappa}),~ x_i<x_l \;,
\label{m_2}
\ee
with a universal power:
\be
\kappa=\frac{(x_l-x_i)d}{d-1}\;,
\label{kappa}
\ee
If $\kappa/z>1$, as argued in the Appendix, the functional form in Eq.(\ref{m_2})
is pure exponential.  This time
dependence is valid for $t<t'_i \sim |\ln m_i|^{z/\kappa}$, for larger
times one goes over to the DG regime with the decay in Eq.(\ref{m_1}).

\subsection{Single spin autocorrelation function}

The nonequilibrium autocorrelation function is defined by
$C(y,t,s)=\langle \sigma_y(t) \sigma_y(s) \rangle$ and in the prepared
state there is no initial magnetization. From the scaling
transformation of the autocorrelation function:
\be
C(y,t,s)=b^{-2x} C(y/b,t/b^z,s/b^z)
\label{scal_C}
\ee
we obtain with $b=t^{1/z}$:
\be
C(y,t,s)=t^{-2x/z} \tilde{C}(y/t^{1/z},s/t)\;.
\label{scal_C1}
\ee
In the following we analyze the scaling function, $\tilde{C}(r,\tau)$,
$\tau \le 1$, in the different limits.

\subsubsection{Bulk behavior}

The scaling function in the bulk, $\lim_{r \to \infty}
\tilde{C}(r,\tau)=\tilde{C}_b(\tau)$, has the asymptotic behavior:
\be
\lim_{\tau \to 0} \tilde{C}_b(\tau) \sim \tau^{(d-x_i-x)/z}\;.
\label{C_b}
\ee
So that for $t \gg s$ we have $C_b(t) \sim t^{-\lambda/z}$, with the
bulk autocorrelation exponent: $\lambda=d-x_i+x$.

\subsubsection{Surface behavior}

As for the magnetization relaxation we have to discriminate between
two regimes: the DG and the CD dynamics, for $x_i>x_l$ and $x_i<x_l$,
respectively.

\paragraph{Domain growth: $x_i > x_l$}

In this case domain growth takes place in the surface region, too.
Therefore for small $r$ we can separate the $r$-dependence from the
scaling function in Eq.(\ref{scal_C1}), leading to:
\be
\lim_{r \to 0}\tilde{C}(r,\tau)=r^{2(x_l-x)} \tilde{C}_l(\tau), \quad x_i>x_l \;.
\label{C_1}
\ee
The small $\tau$ dependence of the scaling function $
\tilde{C}_l(\tau)$ is the same as in Eq.(\ref{C_b}) for the bulk
quantity: $\tilde{C}_l(\tau) \sim \tau^{(d-x_i-x)/z}$. Consequently in
the limit $t \gg s$ we have for the surface autocorrelation function
$C_l(t)\sim t^{-\lambda_l/z}$, with the surface autocorrelation
exponent: $\lambda_l=\lambda+2(x_l-x)=d-x_i-x+2x_l$.\cite{RiCz_95}

\paragraph{Cluster dissolution: $x_i < x_l$}

In this case typically no new domains are formed in the surface
region. The autocorrelation function of the system is therefore
dominated by such rare regions in which at preparation there is an
ordered cluster. The probability of the existence of such a region is
very small as it decreases exponentially with its volume.  The
relaxation time, however, associated to such a cluster is very large,
it is the exponential function of the typical size of an interface in
this cluster. Combining these two effects we arrive to a stretched
exponential dependence, as described in the Appendix. Since the
starting cluster structure does not depend on the waiting time the
surface autocorrelation function for a small $t-s \ll s$ is
stationary. In this case both $r \to 0$ and $1-\tau \to 0$, so that
the appropriate scaling combination is $\rho=(1-\tau)^{1/z}/r$. In
terms of this the surface autocorrelation function in the CD regime is
\be
C_l(t) \sim \exp(-C' (t-s)^{\kappa/z})
\label{C_12}
\ee
for $\kappa/z<1$, with the
exponent $\kappa$ given in Eq.\ (\ref{kappa}), whereas for $\kappa/z>1$ the 
functional form in Eq.\ (\ref{C_12}) is pure exponential.

\subsection{Manifold autocorrelations}

Here we consider a $d'$ dimensional manifold, $\cal{M}$, which is
located at a distance $y$ from the surface. The manifold
order-parameter is the sum of the individual operators, $S_y=\sum_{i
  \epsilon \cal{M}} \sigma_y(i)$, and the manifold autocorrelation
function is defined by $G(y,t,s)=\langle S_y(t) S_y(s) \rangle$. This
quantity, as well as manifold persistence, has been introduced and
studied in the bulk of the system by Majumdar and Bray.\cite{mb03}
First, we consider the equal-time autocorrelation function,
$G(y,t,t)=\langle \sum_{i \epsilon \cal{M}} \sigma_y(i) \sum_{j
  \epsilon \cal{M}} \sigma_y(j) \rangle$, and note that in the
disordered initial state at $t=s=0$ it is zero, since $S_y=0$.  As time
goes on correlations between sites within the correlated domain of
linear size, $\xi \sim t^{1/z}$, are built, so that for a bulk
manifold contribution of terms with alternating signs will lead to $\langle 
\sigma_y(i) \sum_{j \epsilon \cal{M}} \sigma_y(j)
\rangle \sim t^{-2x/z}$ and therefore $G_b(t,t) = \lim_{y \to \infty}
G(y,t,t) \sim t^{(d'-2x)/z}$, provided
$d'>2x$.  If, however, $d'<2x$, since the equal-time autocorrelation
function can not decrease, $G_b(t,t)$ will approach a finite limiting
value within a characteristic (microscopic) time. In this case
$G_b(t,t)$ being independent of the waiting time the manifold
autocorrelation function is stationary, $G_b(t,s)=G_b(t-s)$. This type
of results are in complete agreement with the mean-field and scaling
results of Ref. \onlinecite{mb03}.

The above observations can be cast into a scaling relation for the manifold
autocorrelation function as:
\be
G(y,t,s)=t^{(d'-2x)/z} \tilde{G}(y/t^{1/z},s/t)\;,
\label{scal_G1}
\ee
in which the scaling function, $\tilde{G}(r,\tau)$,  has different analyticity properties
in the different regimes.

\subsubsection{Bulk behavior}

The scaling function in the bulk is denoted as $\lim_{r \to \infty}
\tilde{G}(r,\tau)=\tilde{G}_b(\tau)$.

\paragraph{Non-stationary regime: $d'>2x$}

Here $\tilde{G}_b(\tau)$ is analytic for $\tau \ge 1$ and has the limiting behavior:
\be
\lim_{\tau \to 0} \tilde{G}_b(\tau) \sim \tau^{(d-x_i-x)/z}\;.
\label{G_b}
\ee
So that for $t \gg s$ we have $G_b(t) \sim t^{-\lambda'/z}$, with the
bulk manifold autocorrelation exponent: $\lambda'=d-d'-x_i+x$. Note
that for the global autocorrelation function with $d'=d$ the decay
exponent, $\lambda'/z$, is just the initial slip exponent, $\theta$.

\paragraph{Stationary regime: $d'<2x$}

In this regime the scaling function is non-analytic at $\tau=1$ and given for $1-\tau \ll 1$ as:
\be
\tilde{G}_b(\tau) \sim (1-\tau+\tau_0)^{(d'-2x)/z}\;,
\label{G_b1}
\ee
where $\tau_0$ is related to the microscopic time-scale until which $G_b(t,t)$ reaches its limiting value.
In the limit, $1-\tau \gg \tau_0$, we combine Eq.(\ref{G_b1}) with (\ref{scal_G1}) and
obtain a stationary short-time behavior:
\be
G_b(t,s) \sim (t-s)^{(d'-2x)/z} \;,
\label{scaG_b2}
\ee
in complete agreement with Ref. \onlinecite{mb03}.

\subsubsection{Surface behavior}

Here we consider first the initial value of the manifold
autocorrelations at the waiting time $t=s$, thus we have $\tau=1$. For
equal time operators the short distance expansion is expected to be
the same as for the magnetization in Eq.(\ref{m_1}) so that the
scaling function behaves as:
\be
\lim_{r \to 0} \tilde{G}(r,\tau=1) \sim r^{x_l-x}\;.
\label{G1}
\ee
Consequently the surface manifold autocorrelation function at the
waiting time reads as:
\be
G_l(t,t) \sim t^{(d'-x_l-x)/z}\;.
\label{G2}
\ee
Thus, as for the bulk manifold, the time-dependence of $G_l(t,t)$, and
thus the non-equilibrium dynamics has two different regimes.
For $d'>x_l+x$, when $G_l(t,t)$ is divergent for large $t$ we are in
the standard DG regime. On the contrary for $d'<x_l+x$,
when $G_l(t,t)$ approaches a finite limiting value within a microscopic
time-scale the dynamics for $t>s$ is of the CD type.

\paragraph{Domain growth: $d'>x_l +x$}

For $t>s$ the manifold autocorrelation function has the same short
distance expansion as the single spin autocorrelation function in
Eq.(\ref{C_1}), thus the scaling function, $\tilde{G}(r,\tau)$, for
small $r$ behaves as:
\be
\lim_{r \to 0}\tilde{G}(r,\tau)=r^{2(x_l-x)} \tilde{G}_l(\tau), \quad d'>x_l+x \;.
\label{G_1}
\ee
Note that the form of the short distance expansion is radically
changing for $t>s$ in comparison with that for $t=s$ in Eq.(\ref{G1}).
The scaling function, $\tilde{G}_l(\tau)$, has the same small $\tau$
dependence as its bulk counterpart in Eq.(\ref{G_b}), thus $\lim_{\tau
  \to 0} \tilde{G}_l(\tau) \sim \tau^{(d-x_i-x)/z}$. As a final result
for $t \gg s$ the surface manifold autocorrelation function in the DG
regime behaves as: $G_l(t)\sim t^{-\lambda'_l/z}$, with the surface
manifold autocorrelation exponent:
$\lambda'_l=\lambda'+2(x_l-x)=d-d'-x_i-x+2x_l$.

\paragraph{Cluster dissolution: $d'<x_l +x$}

In this case the correlated sites form isolated clusters and the
dominant contribution of the manifold autocorrelation function is due
to large, but rare clusters. The dilution of the clusters during the
relaxation process for $t>s$ is similar to that of the single spin
autocorrelation function. The only difference is that here the
creation of sites in the domain wall of ordered clusters goes on with
time as $t^{x_l-x}$, which follows from the difference in the short
distance expansions of $G(r,\tau=1)$ in Eq. (\ref {G1}) and that of
$G(r,\tau<1)$ in Eq. (\ref{G_1}). As for the single spin
autocorrelation for short times, $t-s \ll s$, the manifold
autocorrelation function is stationary. This means that as $r \to 0$
and $1-\tau \to 0$ the appropriate scaling combination is
$\rho=(1-\tau)^{1/z}/r$, in terms of which the surface manifold
autocorrelation function is given by
\be
G_l(t) \sim \exp(-C'' (t-s)^{\kappa'/z})
\label{G_12}
\ee
for $0<\kappa'/z<1$
with the exponent $\kappa'=(x_l-x)d/(d-1)$. If $\kappa'/z>1$ then the functional
form in Eq.(\ref{G_12}) is pure exponential.

Here we note that for manifold autocorrelations there is an analogy between
the stationary regime in the bulk and the CD regime at the surface. The difference
in the functional forms in Eqs.(\ref{scaG_b2}) and (\ref{G_12}) is due
to the  relevant inhomogeneity in the latter case: bulk order can only
in a very limited way penetrate into the surface region. Formally, in the
bulk $\kappa'$ should be replaced by zero and therefore the decay is algebraic,
as given in Eq.(\ref{scaG_b2}).

\subsection{Persistence of manifolds}

Persistence of manifolds is given by the probability $P_{pr}(t)$ that
the manifold order-parameter, $\langle S_y(t) \rangle$, has not
changed sign up to time $t$. The bulk manifold persistence has been
analyzed by Majumdar and Bray\cite{mb03} and here we shortly recapitulate
their findings.

\subsubsection{Bulk behavior}

In this case the functional form of persistence can be formally
related to the manifold autocorrelation function at the waiting time,
which according to Eq.(\ref{scal_G1}) is given by $G(t,t) \sim
t^{-\zeta}$, with $\zeta=(2x-d')/z$. In the actual analysis the
theorem by Newell and Rosenblatt \cite{NR} is used in which the long
time decay of the persistence of a Gaussian stationary process is
related to the asymptotic decay of its stationary autocorrelation
function. For $\zeta < 0$ the non-stationary correlator after rescaling
and in $\log t$ variable is transformed to a stationary one. From
this persistence in $t$ is obtained in a  
power-law form, $P_{pr}(t) \sim t^{-\Theta_{pr}}$, with a non-trivial
exponent, $\Theta_{pr}$. For $0<\zeta$ the autocorrelation function in
Eq.(\ref{scaG_b2}) is stationary and one can apply directly the
theorem by Newell and Rosenblatt.\cite{NR}  For $0<\zeta <1$ persistence is in a
stretched exponential form: $P_{pr}(t) \sim \exp(-a t^{\zeta})$,
whereas for $\zeta > 1$ it is pure exponential: $P_{pr}(t) \sim
\exp(-b t)$.

\subsubsection{Surface behavior}

Here we should distinguish between the DG and the CD regimes. In the
DG regime having a nonstationary autocorrelation function persistence can
be analyzed in a similar way as for the analogous regime in the bulk. In
this way we obtain that for $d'>x_l+x$, the surface
manifold persistence is of a power-law form, $P_{pr}(t) \sim
t^{-\Theta'_{pr}}$. In the CD regime, with $d'<x_l+x$, the autocorrelation
function is stationary in the short-time regime, $t-s < s$, but for
longer times it becomes non-stationary. Since the dominant variation
of persistence is related to the stationary part of the correlator it
is natural to assume that persistence has
the same type of stretched or pure exponential time dependence, as the
autocorrelation function. Thus for $0<\kappa'/z<1$, $P_{pr}(t) \sim
\exp(-a t^{\kappa'/z})$, whereas for $\kappa'/z>1$ we have $P_{pr}(t)
\sim \exp(-b t)$.

\section{Numerical results}
We discuss in the following the nonequilibrium dynamical behavior of
various inhomogeneous systems and confront the numerical results with
the predictions coming from the scaling theory just presented. The
models studied include semi-infinite Ising models in two and three
dimensions, the Hilhorst-van Leeuwen model (i.e.  two-dimensional
semi-infinite Ising model with an extended surface defect) and Ising
models with a ladder defect.

\subsection{The models}
\subsubsection{Semi-infinite Ising models}
Static and dynamical properties of critical semi-infinite systems have
been studied very intensively during the last thirty years
\cite{binder,diehl,pleim}. The simplest of these models is the
semi-infinite Ising model defined by the Hamiltonian
\begin{equation}
\label{eq_ham1}
{\mathcal H}=
- \, J_s \sum\limits_{surface} \, \sigma_i \sigma_j - J_b \sum\limits_{bulk} \, \sigma_i \sigma_j
\end{equation}
where the spin located at site $i$ takes on the values $\sigma_i=\pm
1$.  The first sum in Eq.\ (\ref{eq_ham1}) is over nearest neighbor
pairs of surface spins, whereas the second sum is over nearest
neighbor pairs where at least one spin is not located at the surface.
$J_s$ resp.\ $J_b$ is the strength of the surface resp.\ bulk
couplings.

The surface phase diagram of the semi-infinite Ising model is well
established. It is especially simple in two dimensions where surface
and bulk order at the bulk critical temperature $T_c$, independently
of the strength of the surface couplings. In three dimensions,
however, different phase transitions are observed, depending on the
value of $J_s$. For weak surface couplings order can not be maintained
at the surface independently of the bulk. Consequently, surface and
bulk both order at the bulk critical temperature. This transition is
called ordinary transition. For strong surface couplings, however,
with $J_s > r_{sp} \, J_b$ where $r_{sp} \approx 1.50$ for the simple
cubic lattice,\cite{Bin84,Rug92} the surface can order independently
of the bulk at a temperature higher than $T_c$. Reducing the
temperature the bulk than orders at $T_c$ in presence of an already
ordered surface (extraordinary transition). The different critical
lines meet at the multicritical special transition point $r_{sp}$.

In our study of the nonequilibrium critical dynamics in Ising models
with free surfaces we restricted ourselves to the ordinary transition
and to the special transition point, thereby focusing on the
short-time critical dynamics, the dynamical scaling regime encountered
for longer times having been studied recently in Ref.\ 
\onlinecite{pl04} and \onlinecite{cg04}.  Table \ref{table:1} lists
the values of the static and dynamic bulk and surface exponents for
the cases under investigation.  For the ordinary transition we chosed
$J_s=J_b$, but some simulations were also done at other values of
$J_s$ to check for universality.  As usual, the system was initially
prepared in an uncorrelated state and then quenched to the critical
temperature at time $t=0$.  The dynamical evolution of the system was
studied using heat-bath dynamics. In two dimensions systems with
typically $300^2$ spins were simulated, whereas in three dimensions a
typical sample contained $60^3$ spins. Free surfaces were considered
in one direction, whereas periodic boundary conditions were applied in
the remaining directions. No finite-size effects were observed for the
large systems under investigation.

\begin{table}
\caption{Static and dynamic critical quantities of the two- and the three-dimensional Ising models. 
$x$: bulk scaling dimension, 
$x_1$: surface scaling dimension, $x_i$: scaling dimension of the initial magnetization,
$z$: dynamical scaling exponent.
 OT: ordinary transition, SP: special transition point. \label{table:1}}
 \begin{tabular}{|c|c|c|c|c|}  \hline
   & $x$ & $x_1$ & $x_i$ & $z$ \\ \hline
  $d=2$ & $1/8$ & $1/2$ & $0.53$ & $2.17$ \\
  $d=3$, OT & $0.516$ & $1.26$ & $0.74$ & $2.04$ \\
  $d=3$, SP & $0.516$ & $0.376$ & $0.74$ & $2.04$ \\ \hline
  \end{tabular}
  \end{table}

\subsubsection{Hilhorst-van Leeuwen model}
The Hilhorst-van Leeuwen model is a two-dimensional semi-infinite
Ising model with an extended surface defect.\cite{hvl,ipt} We have
studied this model on a square lattice. Whereas the couplings parallel
to the surface have a constant strength $J_1$, the strength of the
couplings perpendicular to the surface vary as a function of the
distance $y$ to the surface:
\begin{equation} \label{eq_hvl}
J_2(y) - J_2(\infty) = \frac{\tilde{A}}{y^\omega}.
\end{equation}
with $\tilde{A}=A \, T_c \sinh (2 J_2(\infty)/T_c)/4$.\cite{Blo83} In
the present work only the marginal case with $\omega =1$ and $J_1 =
J_2(\infty)$ has been considered.  For $A<1$ exact results show that
the local scaling dimension $x_1$ is a continuous function of $A$ with
\begin{equation} \label{eq_hvl2}
x_1 = \frac{1}{2} ( 1- A).
\end{equation}
This remarkable property allows us to study in a very systematic way
the nonequilibrium dynamical behavior, passing from the domain growth
regime to the cluster dissolution regime by only changing the value of
the parameter $A$. Especially, for $A = -0.06$ we have $x_1 = x_i$,
whereas for $A=0$ we recover the pure semi-infinite model with
$x_1=\frac{1}{2}$. Of course, the nonequilibrium bulk quantities $x_i$
and $z$ are left unchanged by the presence of an extended surface
defect.

\subsubsection{Bariev model}
The plane Ising model with a defect line is another inhomogeneous
system displaying non-universal critical behavior. In his work Bariev
\cite{Bar79,ipt} analyzed two types of defect lines: a chain defect,
where a column of perturbed couplings with strength $J_{ch}$ is
considered, and a ladder defect, where modified couplings of strength
$J_l$ connect spins belonging to two neighboring columns (i.e. for
$J_l=0$ the system is separated into two semi-infinite parts).  The
short-time nonequilibrium critical behavior has already been studied
in systems with a chain defect.\cite{sim} In the present work we
focus on Ising systems with a ladder defect.

Bariev's exact results demonstrate the dependence of the
local scaling dimension $x_l$ on the values of the defect coupling.
For the ladder defect he obtains
\begin{equation} \label{bariev1}
x_l = \frac{2}{\pi^2} \arctan^2\left( \kappa_l^{-1} \right)
\end{equation}
with
\begin{equation} \label{bariev2}
\kappa_l = \tanh (\frac{J_l}{T_c})/\tanh (\frac{J}{T_c}).
\end{equation}
$J$ is the strength of the unperturbed interactions (in the following
we set $J=1$), whereas $T_c$ is the critical temperature of the pure
two-dimensional Ising model. Eq.\ (\ref{bariev1}) shows that enhanced
(reduced) defect couplings yield a lower (higher) local scaling
dimension as compared to the perfect two-dimensional Ising model, the
largest value being obtained for the semi-infinite model with $J_l=0$.
We therefore have always $x_l < x_i$ in the case of a ladder defect
and accordingly expect to encounter the domain growth regime for all
values of the defect coupling.

As for the two-dimensional semi-infinite model and the Hilhorst-van
Leeuwen model systems with the linear extent $L=300$ have been studied
typically. For the Bariev model periodic boundary conditions are
considered in all directions.

\subsection{Local magnetization relaxation}
In order to study the relaxation of the magnetization the system is
prepared in an initial state with a small magnetization $m_i$. In the
actual simulations initial magnetizations $m_i = 0.02$ and $0.04$ were
usually considered, for some selected cases other values of $m_i$ were
also investigated.  The data shown in the different figures have been
obtained after averaging over at least 60000 realizations using
different random numbers and different initial states.

The relaxation of the local magnetization is shown in Figures
\ref{Abb1} - \ref{Abb3} for the different models. Recall that
depending on the values of $x_l$ and $x_i$ two different scenarios are
possible.  When cluster dissolution takes place ($x_l > x_i$) scaling
theory predicts a stretched exponential decay (\ref{m_2}) at early
times followed by a power-law decay at later times governed by the
exponent $\theta_l = (x_i-x_l)/z < 0$. When $x_l < x_i$, however, the
local magnetization should display a power-law increase with the
exponent $\theta_l = (x_i-x_l)/z > 0$. Our data completely agree with
this picture.

%%%%%%%%%%%%%%%%%%%%%%%%%%%%%%%%%%%%%%%%%%%FIG 1.%%%%%%%%%%%%%%%%%%%%%%%%%%%%%%%%%%%%%%%%%%%%%%%%%%%%%%
{
\begin{figure}[h]
\centerline{\epsfxsize=3.25in\ \epsfbox{
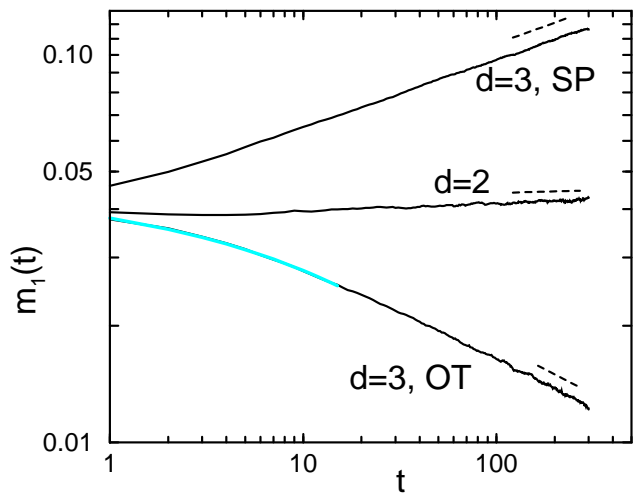}
}
\caption{
  Relaxation of the surface magnetization in the two- and the
  three-dimensional semi-infinite Ising models with $m_i=0.04$. OT:
  ordinary transition, SP: special transition point. The grey line for
  the ordinary transition in three dimensions is a fit to the
  predicted stretched exponential behavior (\ref{m_2}) at early
  times.  The dashed lines indicate the power-law behavior taking
  place in the domain growth regime.  }
\label{Abb1}
\end{figure}
}
%%%%%%%%%%%%%%%%%%%%%%%%%%%%%%%%%%%%%%%%%%%%%%%%%%%%%%%%%%%%%%%%%%%%%%%%%%%%%%%%%%%%%%%%%%%%%%%%%%%%%%%

Looking at Table \ref{table:1} we see that for the three-dimensional
semi-infinite model $x_1 > x_i$ at the ordinary transition, whereas
$x_1 < x_i$ at the special transition point and in the two-dimensional
model.  Accordingly, we expect the cluster dissolution process to take
place only for the ordinary transition in the three-dimensional
system. Figure \ref{Abb1} shows the relaxation of the surface
magnetization for the pure semi-infinite systems. We first remark that
at later times a power-law behavior is observed in all cases. The
values of the exponent extracted in this regime are listed in Table
\ref{table:2} and compared with the expected theoretical value
$\theta_1 = (x_i-x_1)/z$.  In all cases a nice agreement is found
between the numerical value and the theoretical prediction.  At early
times the surface magnetization at the ordinary transition in three
dimensions displays a nonalgebraic behavior. The exponent $\kappa$
obtained from fitting the data with a stretched exponential (grey
line) is listed in Table \ref{table:3}. It nicely agrees with the
scaling prediction (\ref{kappa}).

\begin{table}
\caption{Numerically determined values of the exponent $\theta_l$ governing the power-law behavior of the 
local magnetization when starting from a state with a non-vanishing initial magnetization.
The theoretical prediction is $\theta_l = (x_i-x_l)/z$. 
OT: ordinary transition, SP: special transition point.  \label{table:2}}
\begin{tabular}{|c|c|c|c|}  \hline
\multicolumn{4}{|c|}{pure semi-infinite models} \\ \hline
  & $x_1$ & numerical & theoretical \\ \hline
  $d=2$ &  $1/2$ & $0.015(3)$ & $0.014$ \\
  $d=3$, OT & $1.26$ & $-0.255(4)$ & $-0.255$ \\
  $d=3$, SP & $0.376$ & $0.179(4)$ & $0.178$ \\ \hline
\multicolumn{4}{|c|}{Hilhorst-van Leeuwen model} \\ \hline
$A$  & $x_1$ & numerical & theoretical \\ \hline
$0.75$ & $1/8$ & $0.188(3)$ & $0.187$ \\ 
$0.50$ & $1/4$ & $0.131(3)$ & $0.130$ \\ 
$0.25$ & $3/8$ & $0.076(4)$ & $0.072$ \\
$-0.25$ & $5/8$ & $-0.045(2)$ & $-0.044$ \\
$-0.50$ & $3/4$ & $-0.107(4)$ & $-0.101$ \\ 
$-0.75$ & $7/8$ & $-0.157(3)$ & $-0.159$  \\ 
$-1$ & $1$ & $-0.214(5)$ & $-0.217$ \\ 
$-1.25$ & $9/8$ & $-0.29(1)$ & $-0.274$ \\ 
$-1.50$ & $5/4$ & $-0.37(3)$ & $-0.332$ \\ \hline
\multicolumn{4}{|c|}{Bariev model} \\ \hline
$J_l$  & $x_l$ & numerical & theoretical \\ \hline
0.2 & 0.376 & 0.072(1) & 0.071 \\
0.4 & 0.278 & 0.118(1) & 0.116 \\
0.6 & 0.208 & 0.148(1) & 0.148 \\
0.8 & 0.159 & 0.171(1) & 0.171 \\
1.0 & 0.125 & 0.188(1) & 0.187 \\
1.2 & 0.101 & 0.199(1) & 0.198 \\
1.4 & 0.085 & 0.208(2) & 0.206 \\ \hline
\end{tabular}
\end{table}

\begin{table}
\caption{Numerically determined values of the exponent $\kappa$ governing the stretched exponential behavior
of the cluster dissolution process taking place when $x_l > x_i$. The values have been extracted from (I) 
the relaxation of the local magnetization and (II) from the short-time behavior of the single spin
autocorrelation. The theoretical prediction is given by
Eq. (\ref{kappa}). OT: ordinary transition. \label{table:3}}
\begin{tabular}{|c|c|c|c|c|}  \hline
\multicolumn{5}{|c|}{pure semi-infinite models} \\ \hline
  & $x_1$ & I & II & theoretical \\ \hline
  $d=3$, OT & $1.26$ & 0.81(3) & 0.81(2) & 0.78 \\ \hline
 \multicolumn{5}{|c|}{Hilhorst-van Leeuwen model} \\ \hline
$A$    & $x_1$ & I & II & theoretical \\ \hline
$-0.25$ & $5/8$ & $0.11(4)$ & 0.25(4) & $0.19$ \\ 
$-0.50$ & $3/4$ & $0.45(3)$ & 0.41(3) & $0.44$ \\ 
$-0.75$ & $7/8$ & $0.71(3)$ & 0.65(3) & $0.69$ \\ 
$-1 $ & $1$ & $0.97(4)$ & 0.89(3) & $0.94$ \\ 
$-1.25$ & $9/8$ & $1.17(4)$ & 1.19(3) & $1.19$ \\ 
$-1.50$ & $5/4$ & $1.48(4)$ & 1.54(5) & $1.44$  \\ \hline
\end{tabular}
\end{table}

%%%%%%%%%%%%%%%%%%%%%%%%%%%%%%%%%%%%%%%%%%%FIG 2.%%%%%%%%%%%%%%%%%%%%%%%%%%%%%%%%%%%%%%%%%%%%%%%%%%%%%%
{
\begin{figure}[h]
\centerline{\epsfxsize=3.25in\ \epsfbox{
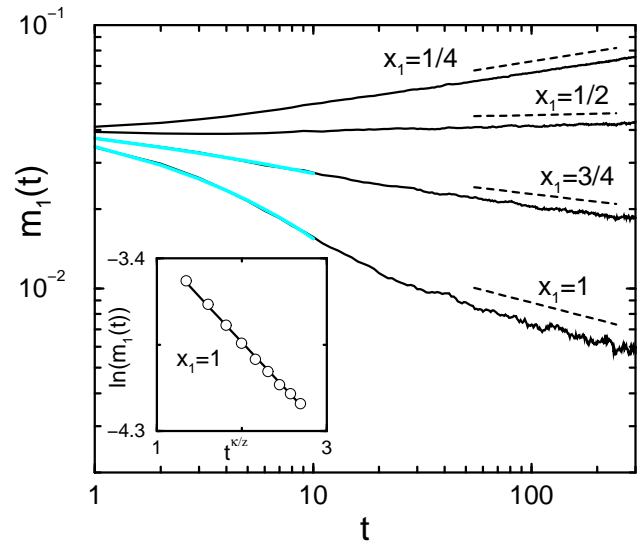}
}
\caption{
  Relaxation of the surface magnetization in the Hilhorst-van Leeuwen
  model for different values of the surface scaling dimension $x_1$.
  The value of the initial magnetization is $m_i=0.04$.  The grey
  lines for the cases $x_1 > x_i \approx 0.53$ are obtained by fitting
  a stretched exponential to the early time data, see Eq.\ 
  (\ref{m_2}).  The expected power-law behavior is indicated by the
  dashed lines.  The inset shows that for $x_1=1$ $\ln(m_1)$ is
  proportional to $t^{\kappa/z}$ at early times, see Eq.\ (\ref{mCD}),
  as expected for the cluster dissolution regime.  }
\label{Abb2}
\end{figure}
}
%%%%%%%%%%%%%%%%%%%%%%%%%%%%%%%%%%%%%%%%%%%%%%%%%%%%%%%%%%%%%%%%%%%%%%%%%%%%%%%%%%%%%%%%%%%%%%%%%%%%%%%

The relaxation behavior of the surface magnetization is shown in
Figure \ref{Abb2} for the Hilhorst-van Leeuwen model with various
values of the scaling dimension $x_1$. Clearly, two different regimes
are observed. For $x_1 < x_i$ the magnetization increases, whereas for
$x_1 > x_i$ it decreases, the decrease being nonalgebraic at early
times. The pure system $x_1 = 1/2$ is very close to the borderline
value 0.53 separating the two cases. This may explain the shallow
minimum observed after the very first time steps before the
magnetization increases again. In all cases a power-law behavior is
encountered at later times, with an exponent that nicely agrees with
the theoretical expectation, see Table \ref{table:2}. The nonalgebraic
decay in the cluster dissolution regime can in all cases be fitted by
a stretched exponential. The values of the exponent $\kappa$ extracted
from these fits are listed in Table \ref{table:3}. As for the
three-dimensional Ising model at the ordinary transition we find a
good agreement with the values obtained from the scaling theory.

Finally, Figure \ref{Abb3} is devoted to the two-dimensional Ising
model with a ladder defect.  As for all defect strengths $J_l$ one has
$x_l < 1/2$, the domain growth regime is the only one accessible
within this model. Indeed, as shown in the Figure, we always observe
an increasing magnetization. Again, the exponent extracted from these
data perfectly agrees with the scaling prediction $\theta_l =
(x_i-x_l)/z$. It has to be noted that the same agreement is obtained
for a chain defect when comparing the theoretical prediction with the
numerically determined values given in Ref.\ \onlinecite{sim}.

%%%%%%%%%%%%%%%%%%%%%%%%%%%%%%%%%%%%%%%%%%%FIG 3.%%%%%%%%%%%%%%%%%%%%%%%%%%%%%%%%%%%%%%%%%%%%%%%%%%%%%%
{
\begin{figure}[h]
\centerline{\epsfxsize=3.25in\ \epsfbox{
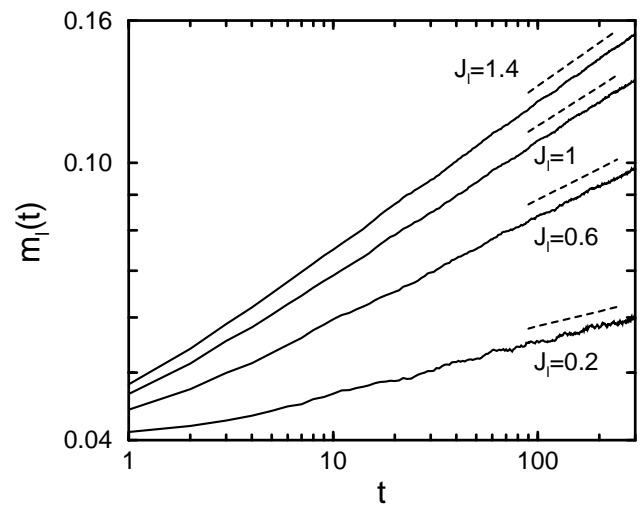}
}
\caption{Relaxation of the surface magnetization in the Bariev model for different strengths
  of the defect coupling $J_l$. The dashed lines indicate the expected
  power-law increase.  }
\label{Abb3}
\end{figure}
}
%%%%%%%%%%%%%%%%%%%%%%%%%%%%%%%%%%%%%%%%%%%%%%%%%%%%%%%%%%%%%%%%%%%%%%%%%%%%%%%%%%%%%%%%%%%%%%%%%%%%%%%

Before turning to the autocorrelation functions, let us briefly
discuss the robustness of the exponents $\theta_l$ and $\kappa$ listed
in Tables \ref{table:2} and \ref{table:3}. For this we investigated
the ordinary transition in the pure semi-infinite systems by changing
the strength of the surface couplings, $J_s$. One then observes that
the value of $\theta_l$ is very robust against modifications of the
surface coupling strength. In all studied cases a power-law behavior
with a constant $\theta_l$ sets in when entering the domain growth
regime.  When dealing with the three-dimensional system, the cluster
dissolution regime is encountered at early times, yielding a stretched
exponential behavior, with a crossover to a power-law at later times.
It is this crossover time which is strongly affected by the strength
of the surface couplings. For weak surface couplings, as for example
for $J_s=J_b/2$, dissolution of clusters is extremely effective in the
surface region, yielding a stretched exponential decay only at the
very first time steps, the system crossing over to the DG regime very
rapidly.  This fact makes the extraction of $\kappa$ in this case very
tedious. On the other hand, when increasing the strength of the
surface interactions, one approaches the special transition point
located at $J_s \approx 1.5 J_b$. As a result complicated crossover
effects are encountered which again make a reliable determination of
$\kappa$ difficult. The value of $\kappa$ reported in Table
\ref{table:3} has been obtained for $J_s=J_b$ where the discussed
effects are only of minor importance. Similar remarks also apply to
the autocorrelation functions discussed in the following.

\subsection{Single spin autocorrelation function}
Following Eq.\ (\ref{C_12}) the CD process should lead to an unusual
stationary behavior of the single spin autocorrelation at early times
where $t-s \ll s$. In our simulations we calculated the local quantity
\begin{equation} \label{c_ising}
C(t,s) = \frac{1}{S} \, \sum\limits_{i \in surface} \langle \sigma_i(t) \, \sigma_i(s) \rangle
\end{equation}
with $t > s$, whereas $S=L^{d-1}$ is the size of the surface or
interface and $L$ is the linear extent of the system. The average is
done over at least 10000 different realizations of the thermal noise.
The sum in (\ref{c_ising}) is over all spins belonging to the surface
or to the defect plane. In the present study we restrict ourselves to
the short-time regime with $t-s \leq 300$.  The dynamical scaling
regime $t$, $s$, $t-s \gg 1$ has been analyzed in Ref.\ 
\onlinecite{pl04}.

Our results for $C(t,s)$ are summarized in Figures \ref{Abb4} and
\ref{Abb5} and in Table \ref{table:3}.  In Figure \ref{Abb4} we plot
$C(t,s)$ versus $t-s$ for the pure semi-infinite cases, whereas the
same is done in Figure \ref{Abb5} in the Hilhorst-van Leeuwen model
for two different values of the scaling dimension $x_1$.  In case of a
stationary autocorrelation at short times the different curves
corresponding to different waiting times should be indistinguishable.
This is exactly what is observed for the cases with a CD regime where
$x_1 > x_i$.  When $x_1 < x_i$, however, the single spin
autocorrelation function is not stationary but depends in a more
complicated way on both times $t$ and $s$ and not merely on the time
difference $t-s$. In the CD regime we again observe a stretched
exponential decay as predicted, see Eq.\ (\ref{C_12}). This is
illustrated in the inset of Figure \ref{Abb4}a where we plot for the
ordinary transition in three dimensions $\ln(C(t,s))$ as a function of
$(t-s)^{\kappa/z}$.  Fitting a stretched exponential function to the
short-time data we can extract the decay exponent $\kappa$. The
resulting values are listed in Table \ref{table:3} together with the
estimates obtained from the relaxation measurements. A good agreement
with the theoretical prediction is observed.

%%%%%%%%%%%%%%%%%%%%%%%%%%%%%%%%%%%%%%%%%%%FIG 4.%%%%%%%%%%%%%%%%%%%%%%%%%%%%%%%%%%%%%%%%%%%%%%%%%%%%%%
{
\begin{figure}[h]
\centerline{\epsfxsize=3.25in\ \epsfbox{
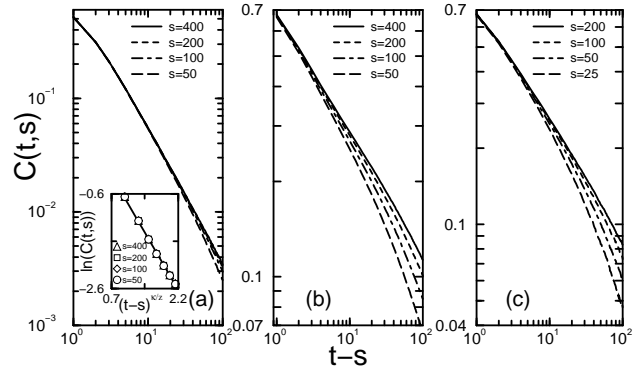}
}
\caption{Single spin autocorrelation vs the time difference $t-s$ for different waiting times $s$ as obtained for the different
  pure semi-infinite systems: (a) ordinary transition and (b) special
  transition point in three dimensions, and (c) ordinary transition in
  two dimensions. When $x_1 > x_i$ a stationary behavior is observed
  with a stretched exponential decay at early times, see inset in (a).
  When $x_1 < x_i$ the autocorrelation depends in a more complicated
  way on both times $t$ and $s$.  }
\label{Abb4}
\end{figure}
}
%%%%%%%%%%%%%%%%%%%%%%%%%%%%%%%%%%%%%%%%%%%%%%%%%%%%%%%%%%%%%%%%%%%%%%%%%%%%%%%%%%%%%%%%%%%%%%%%%%%%%%%

%%%%%%%%%%%%%%%%%%%%%%%%%%%%%%%%%%%%%%%%%%%FIG 5.%%%%%%%%%%%%%%%%%%%%%%%%%%%%%%%%%%%%%%%%%%%%%%%%%%%%%%
{
\begin{figure}[h]
\centerline{\epsfxsize=3.25in\ \epsfbox{
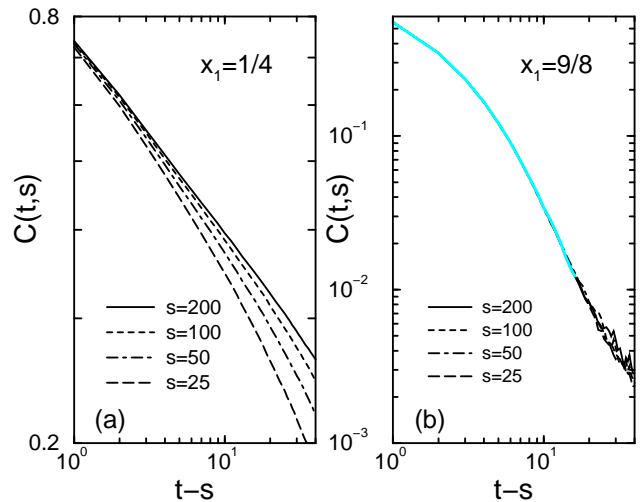}
}
\caption{Single spin autocorrelation as function of the time difference $t-s$ as obtained for the Hilhorst-van Leeuwen
  model with different waiting times. A stationary behavior is
  observed in case (b) where $x_1 > x_i$. The grey line is obtained by
  fitting a stretched exponential to the short time data.  }
\label{Abb5}
\end{figure}
}
%%%%%%%%%%%%%%%%%%%%%%%%%%%%%%%%%%%%%%%%%%%%%%%%%%%%%%%%%%%%%%%%%%%%%%%%%%%%%%%%%%%%%%%%%%%%%%%%%%%%%%%

\subsection{Manifold autocorrelations}
The phenomenological theory presented in Section II yields very
remarkable predictions regarding the surface manifold autocorrelation
function in an inhomogeneous critical system.  For a surface or a
defect plane in two dimensions the surface manifold $\cal{M}$ has the
dimensionality $d'=1$, whereas $d'=2$ for a three-dimensional system.
The surface manifold autocorrelation function is defined by
\begin{equation} \label{G}
G_l(t,s) = \langle S(t) \, S(s) \rangle
\end{equation}
with $S(t) = \sum_{i \in \cal{M}} \sigma_i(t)$, the sum extending over all spins
belonging to the manifold. 

For $t=s$ scaling theory predicts a power-law increase of $G_l(t,t)$
with an exponent $(d'-x_l-x)/z$ whenever $d' > x_l+x$. When $d' <
x_l+x$, however, the surface manifold autocorrelation function should
rapidly tend to a constant. Recalling the possible values of $x_l$ and
$x$ in the models under investigation, we see that for the pure
semi-infinite models and for the Bariev model $G_l(t,t)$ should
display a power-law increase in all cases. Only in the Hilhorst-van
Leeuwen model should one enter the regime with $d' < x_l+x$, with the
amplitude value $A=-0.75$ (i.e.\ $x_1 = 7/8$) separating the two
regimes. This limiting case $x_1 = 7/8$ is shown in Figure \ref{Abb6}a
by the grey line. Indeed, a completely different behavior is observed
for smaller values of $x_1$ (lines above the grey line) when compared
to larger values of $x_1$ (lines below the grey line): in the former
case $G_1(t,t)$ increases with time whereas in the latter case
$G_1(t,t)$ rapidly saturates. In Figure \ref{Abb6}b we compare the data
with $d' > x_l+x$ to the expected power-law increase.  For all studied
cases we find excellent agreement with scaling theory.

%%%%%%%%%%%%%%%%%%%%%%%%%%%%%%%%%%%%%%%%%%%FIG 6.%%%%%%%%%%%%%%%%%%%%%%%%%%%%%%%%%%%%%%%%%%%%%%%%%%%%%%
{
\begin{figure}[h]
\centerline{\epsfxsize=3.25in\ \epsfbox{
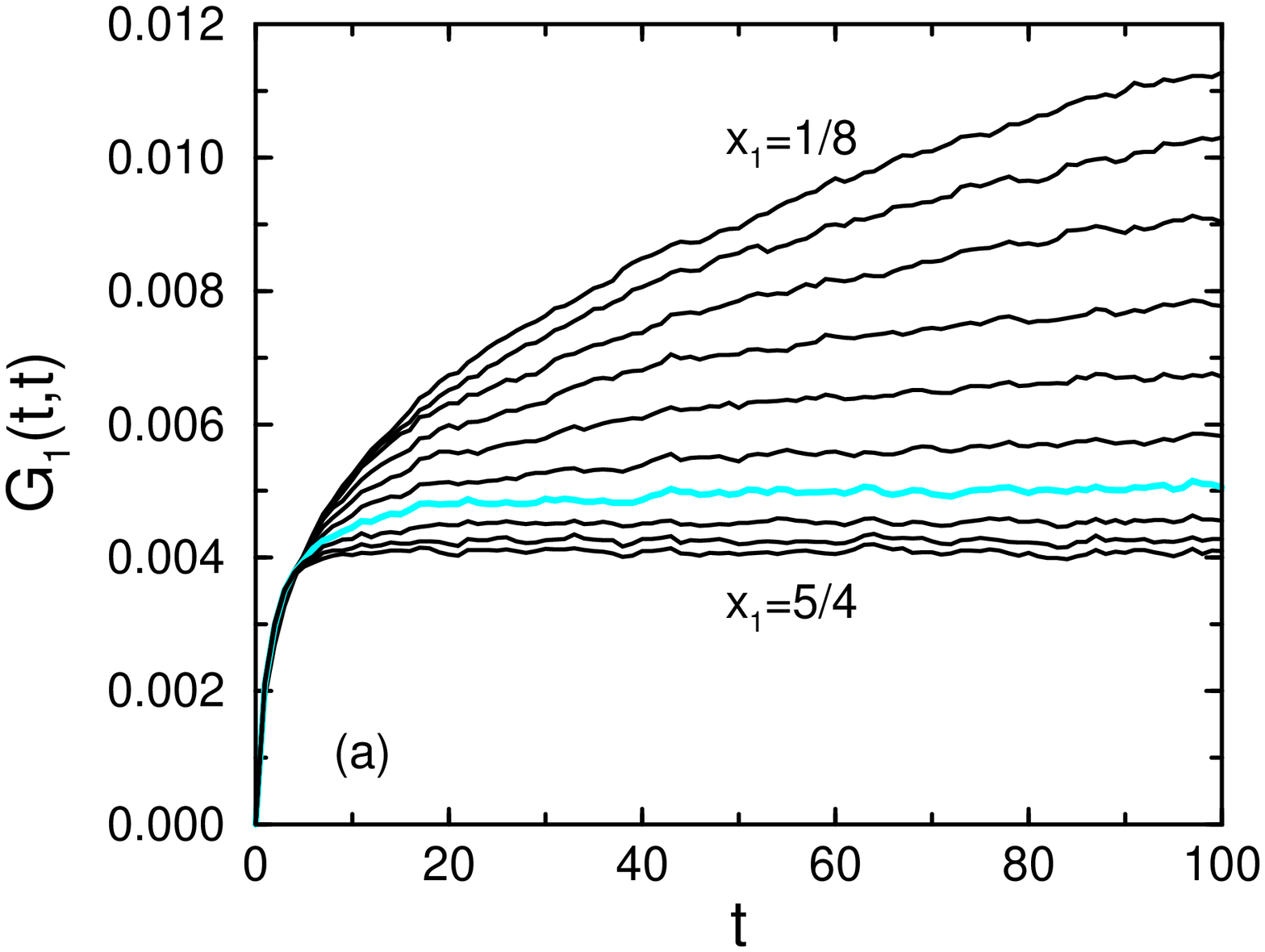}
}
\centerline{\epsfxsize=3.25in\ \epsfbox{
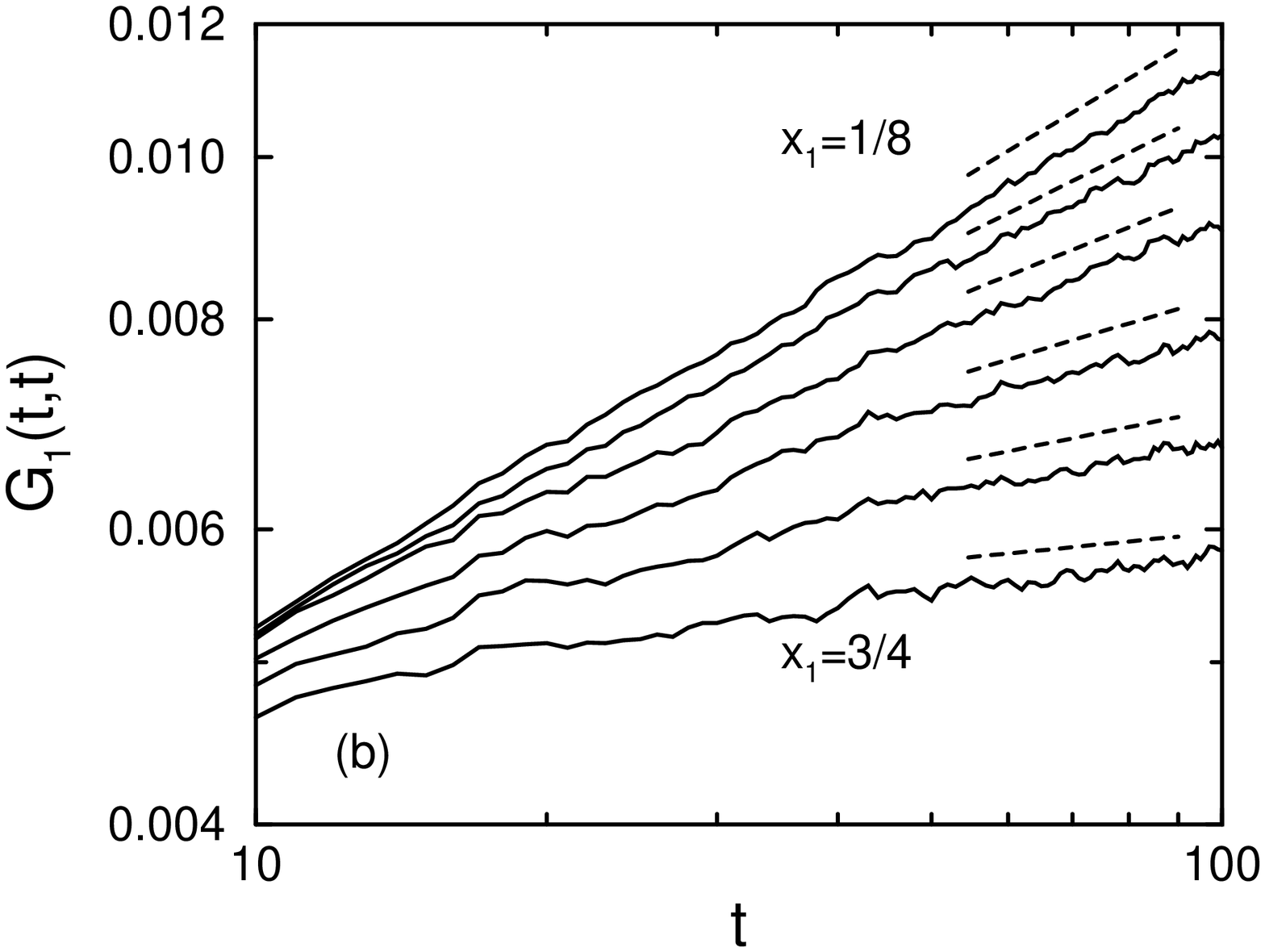}
}
\caption{(a) Surface manifold autocorrelation function for the Hilhorst-van Leeuwen model for various
  values of the scaling dimension: $x_1 = i/8$ with $i=1, \cdots, 10$
  (from top to bottom). The grey line is obtained for $x_1=7/8$ and
  separates the two different regimes with diverging $G_1(t,t)$ for
  $x_1 < 7/8$ and saturating $G_1(t,t)$ for $x_1 \geq 7/8$.  (b)
  Comparison of the increase of $G_1(t,t)$ for $x_1 < 7/8$ with the
  expected power-law (dashes lines) in a log-log plot. From top to
  bottom: $x_1=i/8$ with $i=1, \cdots, 6$.  }
\label{Abb6}
\end{figure}
}
%%%%%%%%%%%%%%%%%%%%%%%%%%%%%%%%%%%%%%%%%%%%%%%%%%%%%%%%%%%%%%%%%%%%%%%%%%%%%%%%%%%%%%%%%%%%%%%%%%%%%%%

As noted in Section II scaling theory predicts for $t>s$ in the DG
regime with $d'>x_l + x$ a change in the form of the short distance
expansion of the surface manifold autocorrelation function as compared
with that for $t=s$. Looking at Eqs.\ (\ref{G2}) and (\ref{G_1}) one
remarks that the scaling combination $s^{(x_l-x)/z} G_1(t,s)/G_1(s,s)$
should then only be a function of $t/s$. As shown in Fig.\ 3 of Ref.\
\onlinecite{pi04} plotting
this scaling combination as a
function of $t/s$ yields a perfect data collapse 
for different values of the local scaling dimension
$x_l$ and for different waiting times, thus
demonstrating the expected $t/s$ dependence.

On the other hand in the CD regime with $d' < x_l +x$ the
autocorrelation $G_1(t,s)$ should just be a function of the time
difference $t-s$, see Eq.\ (\ref{G_12}). This stationary behavior is
indeed observed, as shown in Figure \ref{Abb7} for the Hilhorst-van
Leeuwen model with $x_1 = 9/8$. In the inset we verify the stretched
exponential behavior (\ref{G_12}) by plotting $\ln(G_1)$ as a function
of $(t-s)^{\kappa'/z}$ for $x_1=1$ and $x_1=5/4$.  The values of
$\kappa'$ obtained by fitting the short time data with this kind of
function are gathered in Table \ref{table:4} and compared with the
theoretical expectation $\kappa'=(x_1-x)d/(d-1)$.  For $x_1=5/4$ we
observe a simple exponential, in agreement with $\kappa'/z > 1$.

%%%%%%%%%%%%%%%%%%%%%%%%%%%%%%%%%%%%%%%%%%%FIG 7.%%%%%%%%%%%%%%%%%%%%%%%%%%%%%%%%%%%%%%%%%%%%%%%%%%%%%%
{
\begin{figure}[h]
\centerline{\epsfxsize=3.25in\ \epsfbox{
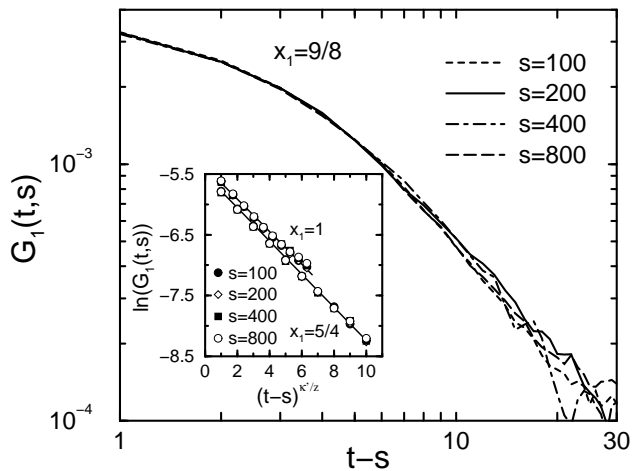}
}
\caption{Early time behavior of the surface manifold autocorrelation function in the CD regime as obtained
  for the Hilhorst-van Leeuwen model with $x_1=9/8$. In the inset the
  stretched (pure) exponential behavior is verified for two different
  values of $x_1$.  }
\label{Abb7}
\end{figure}
}
%%%%%%%%%%%%%%%%%%%%%%%%%%%%%%%%%%%%%%%%%%%%%%%%%%%%%%%%%%%%%%%%%%%%%%%%%%%%%%%%%%%%%%%%%%%%%%%%%%%%%%%

\begin{table}
\caption{Numerically determined values of the exponent $\kappa'$ governing the stretched exponential behavior
of the surface manifold autocorrelation function in the CD regime when $d' < x_l + x$. 
The values have been extracted from (I)
the behavior of surface manifold autocorrelations and (II) from the decay of the persistence probability $P_{pr}$.
The theoretical prediction is given by $\kappa'=(x_1-x)d/(d-1)$. \label{table:4}
}
\begin{tabular}{|c|c|c|c|c|}  \hline
     \multicolumn{5}{|c|}{Hilhorst-van Leeuwen model} \\ \hline
     $A$    & $x_1$ & I & II & theoretical \\ \hline
     $-0.75$ & $7/8$ & $1.41(2)$ & $1.41(2)$ & $1.41$ \\
     $-1 $ & $1$ & $1.63(2)$ & $1.61(2)$ & $1.65$ \\
     $-1.25$ & $9/8$ & $1.86(2)$ & $1.84(2)$ & $1.88$ \\
     $-1.50$ & $5/4$ & $2.03(2)$ & $2.03(2)$ & $2.04$  \\ \hline
     \end{tabular}
     \end{table}

\subsection{Persistence of surface manifold}
Finally, let us discuss the persistence of surface manifolds. The
probability $P_{pr}(t)$ that the manifold magnetization has not
changed sign up to time $t$ is of course directly related to the
functional form of the surface manifold autocorrelation.
Results for the CD regime are shown in Fig.\ 4 of Ref.\ \onlinecite{pi04}.
Plotting $\ln(P_{pr})$ as function of $t^{\kappa'/z}$ we observe straight
lines that indicate that the probability is indeed governed by a stretched
exponential decay whenever $d' < x_l +x$.  Values obtained for
$\kappa'$ nicely agree with those obtained from the surface manifold
autocorrelations, see Table \ref{table:4}. Figure \ref{Abb8} and
Table \ref{table:5} summarize our results for the DG regime with $d' >
x_l +x$. The same behavior is observed in the different models: the
probability $P_{pr}(t)$ decays as a power of time. The values of the
power-law exponent $\Theta_{pr}'$ extracted from the numerical data
are gathered in Table \ref{table:5}. No theoretical expression for
$\Theta_{pr}'$ is available.

\begin{table}[t]
\caption{Numerically determined values of the manifold persistence exponent
$\Theta'$ governing the power-law decay of the persistence probability $P_{pr}(t)$
in the domain growth regime. 
OT: ordinary transition, SP: special transition point. \label{table:5}
}
\begin{tabular}{|c|c|c|}  \hline
\multicolumn{3}{|c|}{pure semi-infinite models} \\ \hline
  & $x_1$ & $\Theta'$ \\ \hline
$d=2$ & $1/2$ & $3.0(2)$ \\
$d=3$, OT & $1.26$ & $4.6(3)$ \\
$d=3$, SP & $0.376$ & $0.61(2)$ \\ \hline
\multicolumn{3}{|c|}{Hilhorst-van Leeuwen model} \\ \hline
$A$    & $x_1$ & $\Theta'$ \\ \hline
$0.75$ & $1/8$ & $0.97(2)$ \\
$0.50$ & $1/4$ & $1.63(3)$ \\
$0.25$ & $3/8$ & $2.0(2)$ \\
$-0.25$ & $5/8$ & $3.9(3)$  \\
$-0.50$ & $3/4$ & $4.5(3)$ \\ \hline
\multicolumn{3}{|c|}{Bariev model} \\ \hline
$J_l$ & $x_l$ & $\Theta'$ \\ \hline
$0.2$ & $0.376$ & $2.3(1)$ \\
$0.4$ & $0.278$ & $1.75(5)$ \\
$0.6$ & $0.208$ & $1.27(3)$ \\
$0.8$ & $0.159$ & $0.90(2)$ \\
$1.0$ & $0.125$ & $0.71(2)$ \\
$1.4$ & $0.085$ & $0.55(1)$ \\
$1.8$ & $0.064$ & $0.40(2)$ \\ \hline
     \end{tabular}
     \end{table}

%%%%%%%%%%%%%%%%%%%%%%%%%%%%%%%%%%%%%%%%%%%FIG 8.%%%%%%%%%%%%%%%%%%%%%%%%%%%%%%%%%%%%%%%%%%%%%%%%%%%%%%
{
\begin{figure}[h]
\centerline{\epsfxsize=3.25in\ \epsfbox{
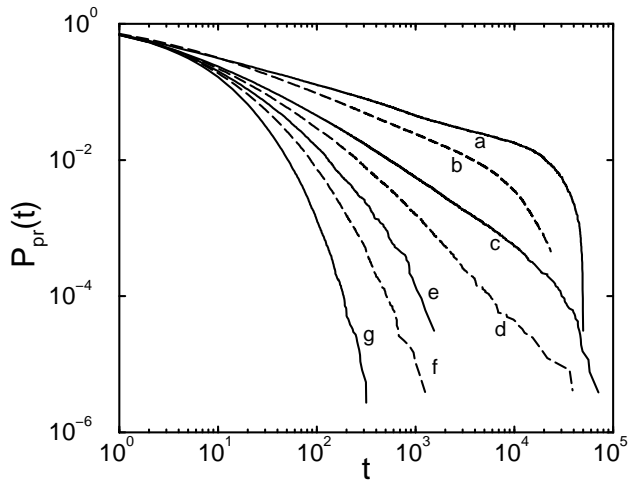}
}
\caption{Manifold persistence probability $P_{pr}(t)$ as a function of $t$ in the domain growth
  regime for various cases.  a: defect line persistence in the Bariev
  model with $J_l=1.8$, b: surface persistence in the
  three-dimensional semi-infinite model at the special transition
  point, c: surface persistence in the Hilhorst-van Leeuwen model with
  $A=0.75$, d: surface persistence in the Hilhorst-van Leeuwen model
  with $A=0.50$, e: defect line persistence in the Bariev model with
  $J_l=0.4$, f: surface persistence in the pure semi-infinite
  two-dimensional Ising model, g: surface persistence in the
  three-dimensional semi-infinite model at the ordinary transition.  }
\label{Abb8}
\end{figure}
}
%%%%%%%%%%%%%%%%%%%%%%%%%%%%%%%%%%%%%%%%%%%%%%%%%%%%%%%%%%%%%%%%%%%%%%%%%%%%%%%%%%%%%%%%%%%%%%%%%%%%%%%

\section{Discussion}

Nonequilibrium critical dynamics involves a time horizon, where the
quench to the critical temperature is made, and this has a long-time,
scale-free effect on the dynamical processes in the system. New
nonequilibrium exponents enter into the theory: the relaxation of the
magnetization and the spin-spin (and manifold) autocorrelations
involve the scaling dimension of the initial magnetization, $x_i$,
whereas persistence of different manifolds involve a new exponent
$\Theta_{\rm pr}$. In a spatially inhomogeneous system, such as at a
free surface or at a defect plane, the local critical behavior is
generally different from that in the bulk and one has to introduce
local critical exponents, such as $x_l$ for the local magnetization.
In this paper we studied nonequilibrium critical phenomena in
inhomogeneous systems, in which the dynamical behavior is the result
of an interplay between a spatial and a temporal inhomogeneity. Our
main result is that in such a system in early times two types of
nonequilibrium processes can take place. Conventional domain growth
phenomena is observed if the temporal surface is more disordered, than
the spatial one. This is the rule in the bulk of the system.  There
is, however, a second, up to now unnoticed process: when the spatial
surface is more disordered then the temporal one, then in early times
local order is reduced and cluster dissolution takes place. This
process is manifested by fast, stretched exponential relaxation and by
stationary, stretched exponential autocorrelations, which, however,
involve a universal exponent.  We have shown by scaling theory and
checked by numerical calculations that as far as relaxation and
autocorrelations are concerned the local nonequilibrium processes can
be fully characterized by the existing exponents, $x_l$, $x_i$ and by
the dynamical exponent, $z$, both in the DG and in the CD regimes.

Our results can be applied or generalized to other types of systems.
We mention that the cluster dissolution scenario has recently been
noticed in nonequilibrium relaxation of reaction-diffusion systems,
such as the contact process.\cite{contact} For another type of
inhomogeneous systems\cite{ipt,pleim} we mention inhomogeneities of
geometrical origin, such as wedges, corners and cones. For example the
local exponent at a corner, $x_c$, depends on the opening
angle,\cite{car83,ipt} $\varphi$, and in two dimensions it is related
to the surface exponent as $x_c=x_1 \pi/\varphi$ through conformal
invariance.\cite{cardy} Generally local order at a corner is weaker
than at a free surface, therefore one expects that the cluster
dissolution scenario often applies for such systems.

\acknowledgements We thank the Regionales Rechenzentrum Erlangen for
the extensive use of the IA32 compute cluster.  This work has been
supported by the Hungarian National Research Fund under grant No OTKA
TO34183, TO37323, TO48721, MO45596 and M36803.

\appendix*
\section{Nonequilibrium dynamics with cluster dissolution}

Nonequilibrium dynamics of inhomogeneous systems, in which the local
(surface) scaling dimension, $x_l$, is larger than the scaling
dimension of the initial magnetization, $x_i$, is governed by cluster
dissolution (CD).\cite{pi04} (Similar scenario holds for the surface
manifold autocorrelation function for $d'<x_l+x$.) In this case the
nonequilibrium growth process, which goes in time as $\sim t^{x_i/z}$,
is much weaker than critical local relaxation, which has a time
dependence of $t^{-x_l/z}$. As a result typically no ordered domains
are created in the surface region and the dynamics is governed by
large correlated clusters which are present with a very small
probability in the initial state.\\

\subsection{Relaxation}
\label{App_rel}
Let us start with an initial state having a magnetization, $m_i$. The
probability of having a cluster of linear size, $l$, is given by $P(l)
\sim m_i^{l^d}$, where $d$ is the dimension of the system. During the
relaxation process the mass of a cluster is diluted by a factor of
$t^{(x_i-x_l)/z}$, and the cluster is dissolved, if a domain wall of
size $l^{d-1}$ is created in it. From the relation $t(l)^{
  (x_i-x_l)/z}l^{d-1}=O(1)$ we obtain for the typical time-scale of a
cluster of linear size, $l$: $t(l) \sim l^{(d-1)z/(x_l-x_i)}$. After
time $t>t(l)$ only those clusters exist, which have an original size
larger than $l$. The magnetization, $m(t)$, is just given by the
contribution of these large clusters: $m(t) \sim \sum_{l>l(t)} P(l)
l^d$. From this we obtain in leading order:
\be
\ln(m(t)/m_i)  \sim -t^{\kappa/z},\quad
\kappa = \frac{(x_l-x_i)d}{(d-1)}\;.
\label{mCD}
\ee
Note, however, that the relaxation can not be faster, than in the {\it non-critical
case}, when it is pure exponential, i.e. the exponent in Eq.(\ref{mCD}) reads as:
$min(\kappa/z,1)$. This result is compatible with the scaling form in Eq.(\ref{m_2}) and
(\ref{kappa}).

\subsection{Single spin autocorrelations}

The reasoning is similar as for relaxation and we repeat that for a
completely uncorrelated initial state with zero magnetization the
probability of having a large cluster of linear size, $l$, is given by
$P(l) \sim \exp(-Al^d)$ and the corresponding relaxation time is
$t_r=t(l) \sim l^{(d-1)z/(x_l-x_i)}$. The autocorrelation function is
obtained by performing an average over the clusters:
\be
C(\tau) \sim \int P(t_r) \exp(-\tau/t_r) {\rm d} t_r  \sim  \exp(-C \tau^{\kappa/z})\;,
\label{auto-}
\ee
in terms of $\tau=t-s < s$. If it happens that $\kappa/z>1$, then the
decay is pure exponential. Note that the autocorrelation function is
stationary for early times and is compatible with Eq.(\ref{C_12}).

\subsection{Manifold autocorrelations}

For surface manifold autocorrelations the regime of cluster
dissolution is valid for $d'<x_l+x$. In this case the dilution of
ordered clusters goes in time as $t^{(x-x_l)/z}$. As a consequence
results for the single spin autocorrelation function can be easily
translated, just in Eq.(\ref{auto-}) the decay exponent $\kappa$ has
to be replaced by $\kappa'=(x_l-x)d/(d-1)$. The corresponding results
are in Eq.(\ref{G_12}).

%\hskip -.5cm

\end{document}